\documentclass[useAMS,usenatbib,,onecolumn]{mn2e}
\usepackage{graphicx}
\usepackage{amsmath}
 \usepackage{color}
 \usepackage[normalem]{ulem}

\title[magnetic-field oscillation model of pulsar]%
{Magnetic-field oscillation model of pulsar radio emission: prediction of an observable effect}

\author[Z.-X. Liang, \& Y. Liang]%
{Zhu-Xing Liang$^{1}$\thanks{zx.liang55@hotmail.com (LZX); liang.y.jlu@hotmail.com (LY)}
and Yi Liang$^{2\star}$ \\
$^{1}$18-4-102 Shuixiehuadu, Zhufengdajie, Shijiazhuang, Hebei 050035, China\\
$^{2}$55-1-302 Shuixiehuadu, Zhufengdajie, Shijiazhuang, Hebei 050035, China}
\begin{document}

\date{}

\pagerange{\pageref{firstpage}--\pageref{lastpage}} \pubyear{2002}

\maketitle

\label{firstpage}

\begin{abstract}
Since the discovery of pulsars, the rotating-lighthouse model has been the choice of model to explain the radiation pulsation of pulsars. After discovering that some main sequence stars (e.g., CU Virginis) and ultracool dwarfs (e.g., TVLM 513-46546) also behave like pulsars, the lighthouse model was again adopted to explain their pulse signals. Our research found that if we use the magnetic-field oscillation (MO) model, we can explain the pulse radiation results better regardless of whether its source is a neutron star, a main-sequence star or an ultracool dwarf. We propose a verifiable prediction that can be used to evaluate the MO model. Our prediction is that there is a 90$\degr$ phase lag between the magnetic field and radio signal of TVLM 513-46546, and the zero-crossing point of the magnetic field is the moment when the direction of the light's circular polarization is reversed. No new observations are needed to check this prediction, but certain existing data needs to be re-mined.

\end{abstract}

\begin{keywords}
polarization -- methods: data analysis -- pulsars: general -- stars: magnetic fields.
\end{keywords}

\section{Introduction}

In 1967, Jocelyn Bell and Antony Hewish discovered the first pulsar. Shortly thereafter, the rotating-lighthouse model was employed to explain the periodic radiation of pulsars
\citep{Gold68,Radhakrishnan69}. To date, the lighthouse model has become the only trusted model. This model suggests that a pulsar is a rapidly rotating, highly magnetized neutron star, and the magnetic poles can emit beams with high directionality. These beams of light sweep through the vast universe like beams from a rotating lighthouse; the pulsar is seen when the beam sweeps across Earth.

To produce a highly directional beam, an extreme magnetic field, as high as $10^{8}$ -- $10^{12}$ Gauss is required around the neutron star \citep{Lyne06}. However, observations in the past 20 years found that some non-compact stars with weak magnetic fields (thousands of Gauss) can also emit periodic radiation like pulsars; they include the AP star CU Virginis \citep{Leto06, Lo12}, the BP star HD 133880 \citep{Das18}, the ultracool dwarfs TVLM 513-46546 \citep{Hallinan07}, LSR 1835+3259 \citep{Hallinan08}, 2MASSW J0746425+200032 \citep{Berger09}, 2MASS J13142039+130011 \citep{McLean11}, 2MASS J00361617+1821104 \citep{Berger05} and WISE J062309.94-045624.6 \citep{Rose23}. Except for the difference in the timescale of the pulse period, there is no significant difference between the radiation signals of these star systems and that of pulsars. For convenience in later discussions, we call these objects that emit periodic signals like pulsars but are not neutron stars `quasi-pulsars'.

Although the pulse periods of quasi-pulsars are several hours long, their pulse profiles, duty cycles, and coherence characteristics are remarkably like those of pulsars. Given these similarities, several authors have employed the lighthouse model to explain their pulse signal. Meanwhile, they consider the underlying radiation mechanism of quasi-pulsars to be coherent electron cyclotron maser emission \citep{Trigilio08, Hallinan07}.

We previously published a method that can test the lighthouse model \citep{Liang14}. Using our method, \citet{Pol18} reported a result that demonstrated the validity of the lighthouse model. However, when re-examining their routine, we found that the pulse frequency of PSR J0737-3039B used by them was 0.3605 Hz, reaching a cumulative error of 0.0039 Hz after 64 accumulations of this low-precision frequency value. This error has far exceeded the orbital frequency of the binary system (0.000113 Hz), making the calculation results meaningless. After replacing the low-precision value with a high-precision value (0.36056035506 Hz), we could not reconstruct their results. Therefore, to confirm the lighthouse model, a reassessment is needed. We also found that even using a high-precision frequency value, the three frequency points (corresponding to prograde rotation, no rotation, or retrograde rotation) still deviated from the extreme value (maximum) of the power spectrum, thus making it impossible to draw reliable conclusions. There are two possible reasons: first, the available data spanned only 344 s long, whereas the orbital period was as long as 8820 s. To extract the influence of the orbital frequency, the amount of data may be insufficient. Second, a centroid correction was required in the calculation, and therefore the correction procedure may have generated systematic errors. The exact cause remains to be further investigated by experts.

We have proposed a magnetic-field oscillation model (MO model) \citep{Liang07} as a competing model to the lighthouse model. Although the MO model provides concise explanations for most of the radiation features of pulsars, it was never recognized in the astrophysics community. The emergence of quasi-pulsars provides new evidence supporting the MO model. Quasi-pulsars not only provide radio pulse signals but also magnetic field strength signals. The correlation between these two periodic signals provides additional information for our study, leading to some advancements. Here, we introduce our new findings and propose a verifiable prediction for testing the MO model.

\section{Introduction to the MO Model}

Historically, the Schuster-Wilson hypothesis \citep{Schuster11, Wilson23} proposed that if the rotational speed of a star is high enough, the star can generate and maintain a magnetic field. However, this hypothesis was refuted by the anti-dynamo theorem and an experiment \citep{Cowling33, Blackett52}. Our analysis found that both Cowling's theorem and Blackett's experiment negated stable magnetic fields only, but did not negate alternating magnetic fields. If the requirement of a stable magnetic field is replaced by that of an alternating magnetic field, the Schuster-Wilson hypothesis may be resurrected.

A recognized fact is that the oscillation period of the solar magnetic field is 22 years. Regardless of its mass, if the Sun were to collapse and become a neutron star, during the collapse, its magnetic-field oscillation may not stop but instead quicken. Its period may shorten to seconds or even milliseconds, and finally become a pulsar. This is the genesis of the MO model.

The MO model uses four basic assumptions:
\begin{enumerate}
  \item \textbf{The flip oscillation is a common attribute of the magnetic fields of all stars.} First, the solar magnetic field exhibits oscillatory behavior. Second, \cite{Babcock51, Babcock58} reported several stars with a magnetic period of approximately one week. After a thorough study of those stars, he hypothesized that their magnetic fields undergo oscillations. On these observations, we believe that flipping oscillations may be a common attribute of the magnetic fields of all stars.
  \item \textbf{The oscillation frequency of the magnetic field is positively correlated with its rotational frequency.} Neutron stars are formed by the collapse of main sequence stars. With the conservation of angular momentum, their rotational speeds are inevitably much higher than that of the Sun. Their pulse frequencies (pulse arrival rate) are also much higher than the oscillation frequency of the Sun's magnetic field. Therefore, we assume that the oscillation frequency is positively correlated with the rotational frequency, but they are no longer equal.

  \item \textbf{The strength of the stellar magnetic field is positively correlated with its oscillation frequency.} The reported pulse frequencies of the quasi-pulsars are higher than the oscillation frequency of the solar magnetic field, and their magnetic field strengths are also higher than that of the Sun. This may be a universal law.
  \item \textbf{A magnetic field oscillation induces a oscillating electric field, which, at its peak, produces a coherent radiation pulse.} The higher the induced electric field, the higher the electron acceleration; therefore, coherent radiation is more favored to occur.

\end{enumerate}

When a dipole magnetic field oscillates, a doughnut-shaped radiation region is formed parallel to the equatorial plane. In the radiation region, an alternating electric field is established around the star. This electric field accelerates electrons. At every location at the same altitude, the electric field strength is the same, therefore, the highly ordered distribution of magnetic field, electric field, electron speed, and electron acceleration creates the best conditions for coherent radiation.

Based on the second and third assumptions above, the induced electric field is negatively correlated with the period of oscillation \emph{P}. Therefore, both the acceleration and speed of the electrons are also negatively correlated with \emph{P}. That is, quasi-pulsars feature long periods and low electron speeds, whereas pulsars exhibit short periods and high electron speeds.

According to current particle radiation theory, cyclotron radiation occurs with low electron speeds whereas synchrotron radiation occurs with high electron speeds. Therefore, quasi-pulsars with lower electron speeds produce cyclotron radiation, whereas pulsars produce synchrotron radiation.

\subsection{Cyclotron radiation of quasi-pulsars}

Without considering coherent radiation, the characteristic of cyclotron radiation is that the radiation parallel to the magnetic axis is fully circularly polarized whereas the radiation perpendicular to the magnetic axis is fully linear polarized, with the circular-polarized light being twice as strong as the linear-polarized light. In the MO model, the notion of magnetic inclination angle is not needed, and the rotation axis is always aligned with the magnetic axis. The equatorial plane where the pulse radiation occurs is like a huge planar radar antenna. This antenna lets coherent light concentrate along the rotation axis and forms a light beam with high directionality. If our line-of-sight coincides with the rotation axis, we see very bright circular polarized light. In this process, the coherence property is of great significance for the beam's directionality.

If our line-of-sight is parallel to the equatorial plane, we see fully linear-polarized light. However, the radiation near the equatorial plane is only energy that leaks around the antenna and is not directional. The radiation flux is dispersed uniformly in all directions, encompassing a full 360$\degr$. Hence, only a small fraction of light is emitted towards Earth. As a result, only circular-polarized light is easily observed from quasi-pulsars.

\subsection{Synchrotron radiation of pulsars}

%%%%%%%%%%%%%%%%%%%%%%%%%%%%%%
\begin{figure}
\includegraphics[width=160mm]{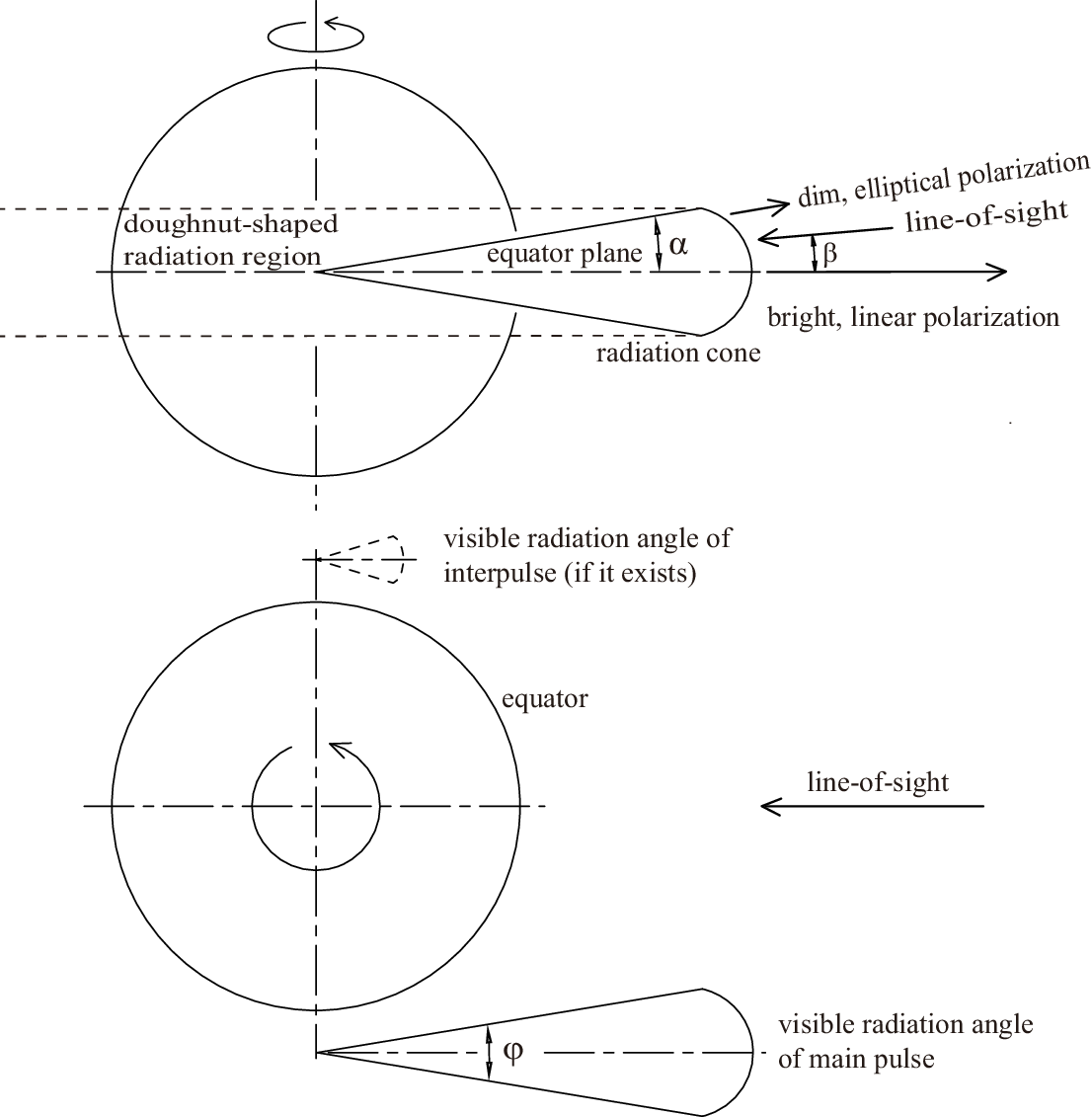}
\caption{Polarization and brightness from synchrotron radiation under the MO model. The radiation region is toroidal in shape parallel to the equatorial plane. Only the flux inside the radiation cone can reach Earth. The radiation cone does not co-rotate with the neutron star.}
\label{fig:fig1}
\end{figure}
%%%%%%%%%%%%%%%%%%%%%%%%%%%%%%

According to the MO model and synchrotron radiation theory, the radiation region is toroidal in shape lying symmetrically to the equatorial plane. The radiation flux is concentrated in a circular space with a field angle of 2$\alpha$ (Fig. \ref{fig:fig1}); the axial flux disappears.

In the MO model, the trajectory plane of the electron motion is always parallel to the equatorial plane. With this premise, all the observational effects listed below can be inferred from the theory of synchrotron radiation.
\begin{enumerate}
  \item The centre of the radiation cone has the highest flux density, whereas that of the edge is lowest.
  \item When the line-of-sight is parallel to the equatorial plane, fully linear-polarized light is observed.
  \item When the line-of-sight deviates from the equatorial plane, namely $\alpha > \beta \neq 0$, elliptical-polarized light is observed, here, $\beta$  is the angle between the line of sight and the equatorial plane, as shown in Figure 1. The larger the angle $\beta$, the higher the circular polarization fraction and the lower the flux observed.

      Based on the above three points, a statistical law can be derived: brighter pulsars exhibit a higher degree of linear polarization.
  \item Only a part of the radiation region (the visible radiation region) contributes to the observed flux; the remainder propagates in other directions. The ratio of the visible radiation region and the entire radiation region equals $\varphi/360$, where $\varphi$ is the visible radiation angle in Fig. \ref{fig:fig1}.
  \item The higher the electron speed, the smaller the half-cone angle $\alpha$ and the visible radiation angle $\varphi$.
  \item The alternating electric field results in an alternating electron speed, and hence the angles $\alpha$  and $\varphi$ expand and contract with period \emph{P}. If $\beta \neq 0$, what we see is elliptically polarized light. Because the angle $\alpha$ determines all polarization characteristics, including the polarization position angle (PA), when $\alpha$ changes, PA also changes. The reversal of the magnetic-field direction may be the reason for the jump of PA.
  \item  The hollow radiation cone described in the lighthouse model can be explained with the MO model: a dip in the synchrotron radiation occurs when the magnetic field crosses zero.
  \item The fluctuation in the density of electrons in the visible radiation region creates the complexity in the pulsar signal. Because the rotation period is not equal to the magnetic period, and the electron density in the visible radiation region may fluctuate with the rotation period, the coupling between the two periods may cause pulse null, drift, and even quasi-period patterns (beat frequency).
\end{enumerate}

\section{USING THE MO MODEL TO EXPLAIN THE POLARIZATION CHARACTERISTICS}

Each pulsar and quasi-pulsar have distinct linear and circular polarization fractions. To study the relative contributions of these polarization fractions, \citet{Oswald23} introduced a helpful metric termed the `circular contribution', $\theta$. According to their definition, $\theta$ is the magnitude of the angle of latitude on the Poincar\'{e} sphere. Fully linear polarization corresponds to $\theta$ = 0 whereas fully circular polarization corresponds to $\theta = 90\degr$.

\begin{figure}
\includegraphics[width=160mm]{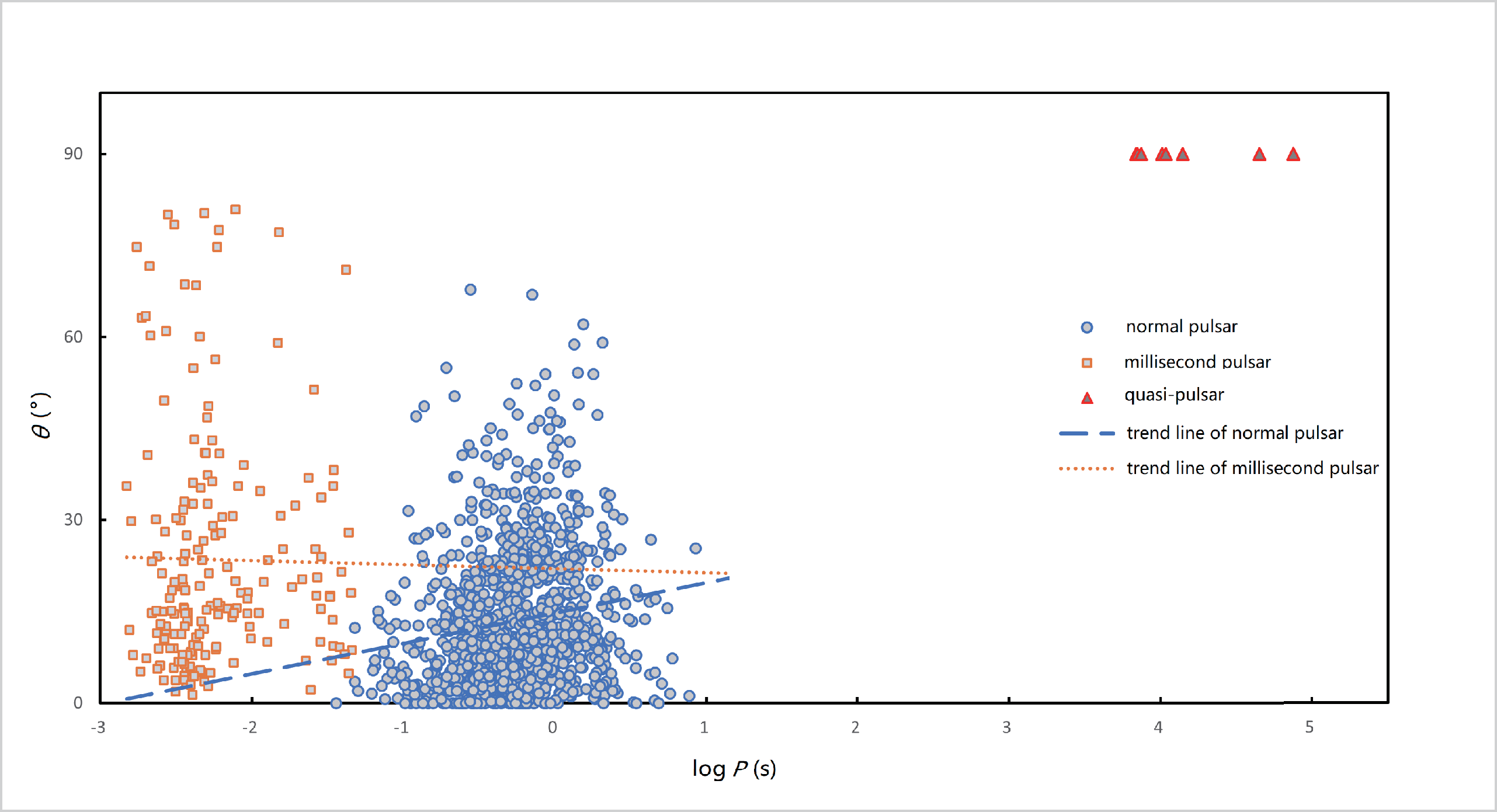}
\caption{Correlation between circular contribution $\theta$ and pulse period \emph{P}.}
\label{fig:fig2}
\end{figure}

\citet{Oswald23} investigated the relationship between $\theta$ and radio frequency. However, we consider the relationship between $\theta$ and pulse period \emph{P} is also noteworthy. If the MO model is correct, based on the analysis above, $\theta$ should be positively correlated statistically with \emph{P}; specifically, the longer the pulse period, the larger the circular contribution $\theta$. To verify this relationship, we plotted the available data of pulsar and quasi-pulsar in  Fig. \ref{fig:fig2}. The data from eight quasi- pulsars are from relational references cited in the Introduction. The data of 1097 normal pulsars are taken from \citet{Oswald23} and \citet{Posselt23}; any overlapping data between the two sources were merged. The data of 189 millisecond pulsars are taken from \citet{Spiewak22}. The periods \emph{P} were obtained from the ATNF pulsar catalogue (ATNF https://www.atnf.csiro.au/research/pulsar/psrcat) \citep{Manchester05} and most of the $\theta$  values are calculated from the definition:
\begin{equation}\label{*}
  \theta = \arctan \frac{|V|}{L}
\end{equation}
where $|V|$ is the absolute circular polarization fraction, $L$ is the linear polarization fraction.
\\

From Fig. \ref{fig:fig2}, we note four observations:
\begin{enumerate}
  \item All quasi-pulsars exhibit fully circular polarization, with no exceptions so far. All of them thus lie in the upper-right corner region.
  \item The overall trend of the whole graph is that $\theta$ is positively correlated with \emph{P}.
  \item Within the normal pulsar population, $\theta$ is positively correlated with \emph{P}. This result is in line with our expectations. Based on the third assumption in Section 2, we make the following inference: period long $\rightarrow$ magnetic field weak $\rightarrow$ induced electric field weak $\rightarrow$ electron speed low. Then according to the theory of synchrotron radiation, electron speed low $\rightarrow$ $\alpha$ large $\rightarrow$ circularly polarized fraction high $\rightarrow$ $\theta$ large. Thus, we expect a positive correlation between $\theta$ and \emph{P}.
  \item Millisecond pulsars are special, possibly because they are not driven purely by rotational energy; accretion (or past accretion) may have had an impact on the $\theta-P$ relationship.
\end{enumerate}

Roughly speaking, the pulsars are indeed in the lower-left corner of Fig. \ref{fig:fig2} and the $\theta-P$ relationship is consistent with the expectations of the MO model.

\begin{figure}
\includegraphics[width=160mm]{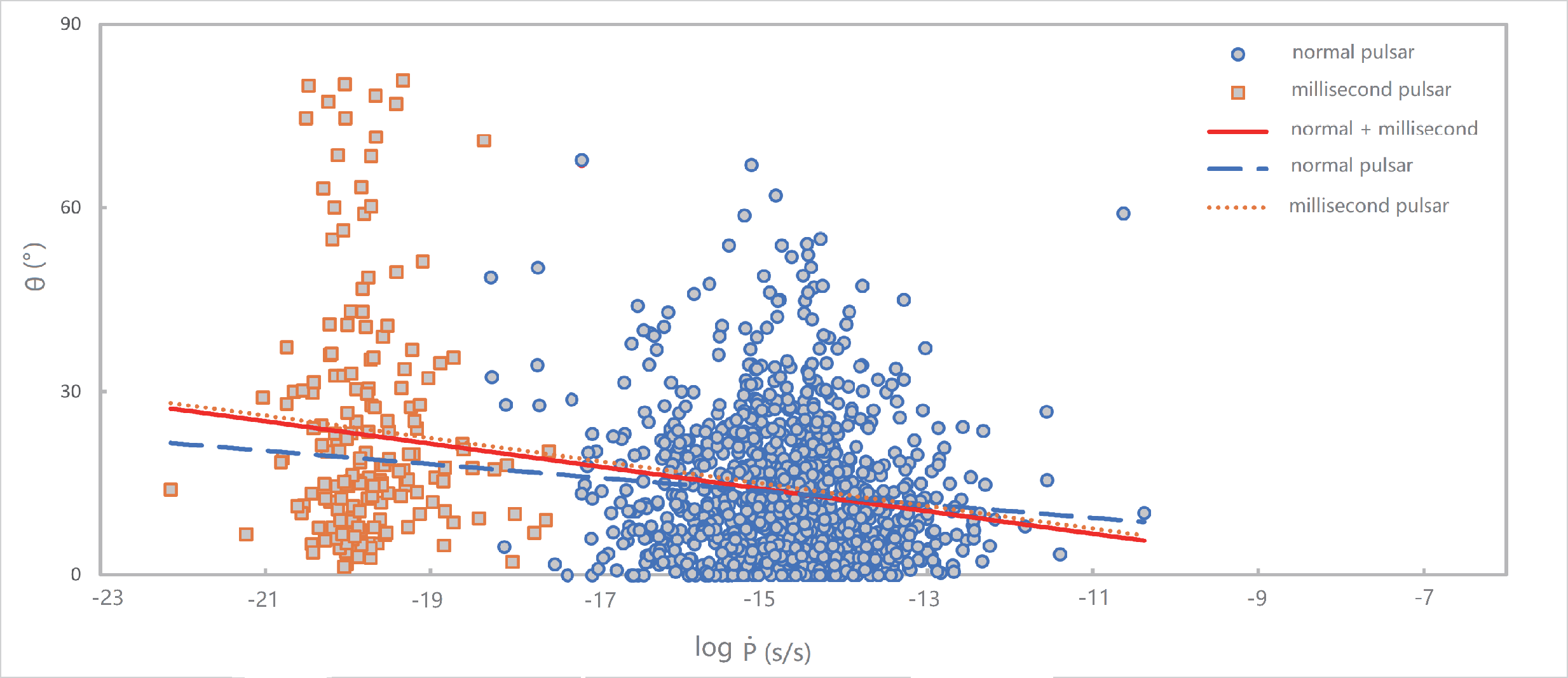}

\caption{Correlation between circular contribution $\theta$ and the derivative of period $\dot{P}$.}
\label{fig:fig3}
\end{figure}

The relationship between $\theta$ and the rate of change in period $\dot{P}$ was plotted in Fig. \ref{fig:fig3}.

Although $\dot{P}$ and spin-down energy $\dot{E}$ are derived from the lighthouse model, they are also important parameters in the MO model. The lighthouse model and the MO model are based on the same assumption: the energy of electromagnetic radiation comes from the rotational energy.

Under the lighthouse model, $\dot{E} = 4\pi^2 I\dot{P}P^{-3}$ (Posselt et al. 2023); here, \emph{P} is the rotation period and \emph{I} the estimated moment of inertia.

In the MO model, \emph{P} is the electromagnetic oscillation period rather than the rotation period, and the rotation period is unknown. Therefore, we cannot directly calculate $\dot{E}$ using the above formula. If we use a positive correlation coefficient to convert the magnetic period into a rotational period (according to the second assumption in Section 2), we can calculate $\dot{E}$ using the same formula again.

In both models, $\dot{P}$ and $\dot{E}$ are related to the power of energy conversion (the rotational energy is converted into electromagnetic radiation energy). High $\dot{P}$ and $\dot{E}$ mean that the energy and speed of the electrons are also high. According to the analysis in Section 2.2, the higher the electron speeds, the lower the value of $\theta$. The justification is: high $\dot{P}$ $\rightarrow$  high $\dot{E}$ $\rightarrow$  high power of energy conversion $\rightarrow$  high speed of electrons $\rightarrow$  low $\theta$, which simply answers the question why is $\theta$ negatively correlated with $\dot{P}$ as shown in Fig. \ref{fig:fig3}.

The MO model clearly explains the relationships (Figs. \ref{fig:fig2} and \ref{fig:fig3}), which should be considered as evidence in its favor. Although this result cannot completely refute the lighthouse model, the explanation provided by the MO model is simple and unmatched by the lighthouse model.

\citet{Desvignes19} provide evidence for the lighthouse model. However, their results simultaneously provide evidence for the MO model. We infer from Fig. \ref{fig:fig1} that in a binary system, if the rotation axis of the pulsar precesses, the radiation cone may gradually leave our line-of-sight and disappear. Before disappearing, the circular polarization fraction inevitably and gradually strengthens, whereas the linear polarization gradually weakens. The results of \citet{Desvignes19} help to confirm this trend (see panels A and C in Fig. 4 in their paper). Those two panels show the evolution of both linear and circular polarizations before the disappearance of the main pulse (MP) of PSR J1906+0746. They perfectly confirmed the inference of the MO model and provided support for the MO model.

The trend lines in Figs. \ref{fig:fig2} and \ref{fig:fig3} are linear fits. For the normal pulsars, the correlation coefficient r between $\theta$ and \emph{P} is 0.16, and that between $\theta$ and $\dot{P}$ is -0.1. Two weak correlations imply that the influence factors of $\theta$ may be very complex. Perhaps, factors such as the radius of the star, temperature, and charge density can all affect the speed of electrons, thereby affecting $\theta$ and the correlation coefficients.

%%%%%%%%%%%%%%%%%%%%%%%%%%%%
%%%%%%%%%%%%%%%%%%%%%%%%%%%%
%%%%%%%%%%%%%%%%%%%%%%%%%%%%
%%%%%%%%%%%%%%%%%%%%%%%%%%%%

\section{A PREDICTION: THERE IS A $90\degr$ PHASE LAG BETWEEN THE RADIO AND MAGNETIC FIELD SIGNALS OF TVLM 513-46546}
%%%%%%%%%%%%%%%%%%%%%%%%%%%%
\begin{figure}
\includegraphics[width=160mm]{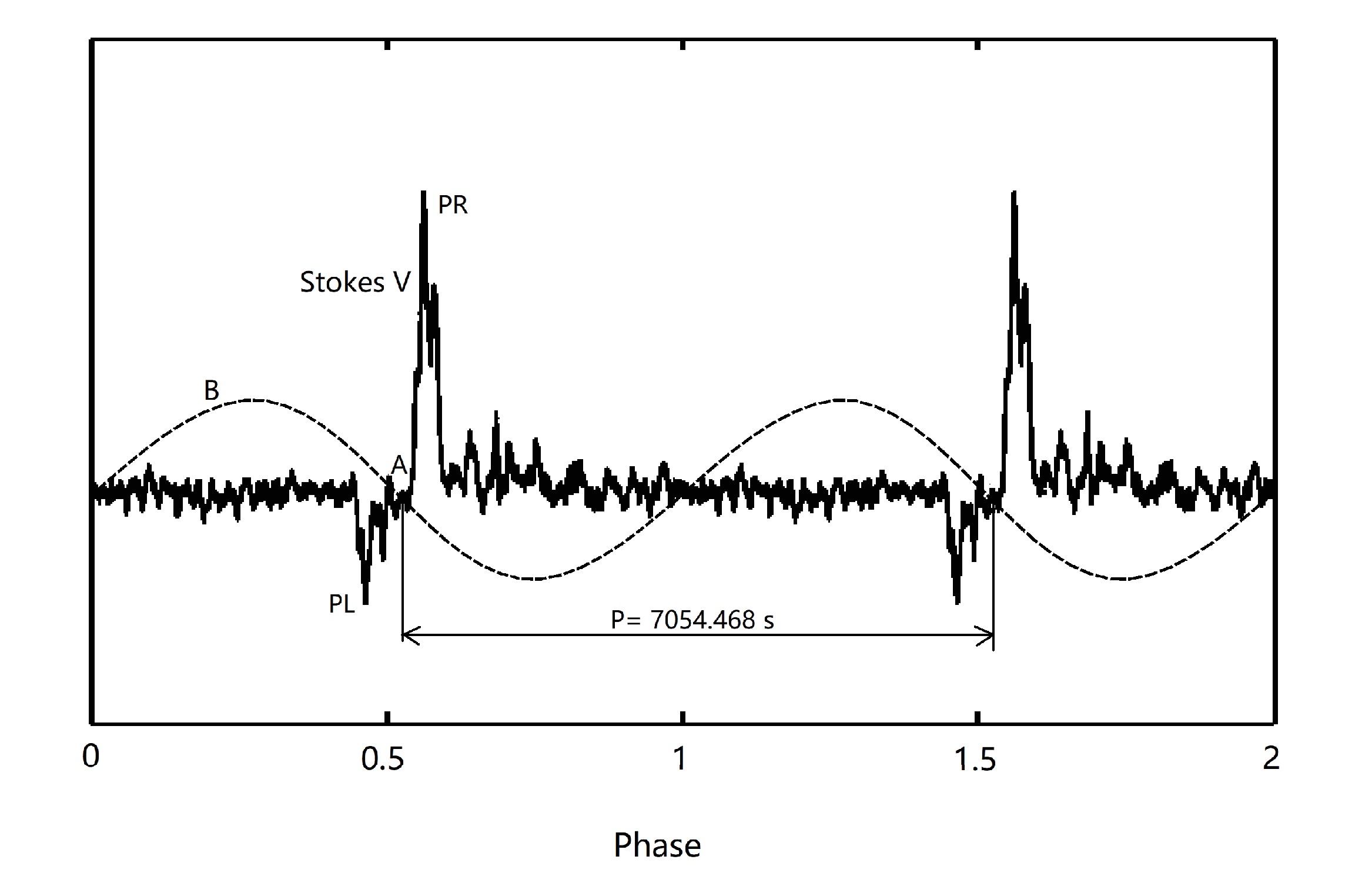}
\caption{Phase relationship predicted by us and based on the MO model.}
\label{fig:fig4}
\end{figure}
%%%%%%%%%%%%%%%%%%%%%%%%%%%%

If the oscillating magnetic field of a star is sinusoidal, the induced electric field and radio radiation reach their peaks at the zero-crossing point of the magnetic field. Therefore, the radio signal lags the magnetic field by a $90\degr$ phase; this phase lag is an important feature of the MO model. The MO model was proposed in 2004 and first made public at the $36^{th}$ COSPAR scientific assembly in Beijing in July 2006. At the time, we gave no attention to reports on quasi-pulsars. Later, we found that three quasi-pulsars indeed exhibited a $90\degr$ phase lag and therefore provided potent support for the MO model. The three quasi-pulsars were CU Virginis \citep{Leto06}, 2MASSW J0746425+200032 \citep{Berger09}, and HD 133880 \citep{Das18}.

Do all quasi-pulsars show a $90\degr$ phase lag? This requires more investigation. We suggest that the investigation starts with TVLM 513-46546 because this does not require new observations. \citet{Hallinan07} observed periodic radio radiation signals from TVLM 513-46546. \citet{Berger08} acquired its periodic changes in magnetic field, and \citet{Wolszczan14} determined its period, $P = 7054.468 \pm 0.007$ s. The only remaining task is to extrapolate the magnetic field curve back to approximately 335 days and then use period \emph{P} to superimpose the magnetic and radio curves (as in Fig. \ref{fig:fig4}).

Based on the MO model, we predict that point A in Fig. \ref{fig:fig4} is both the zero point of the magnetic field and the point where the direction of radio circular polarization is reversed.

If this prediction is confirmed, the $90\degr$ phase lag may be a universal law of quasi-pulsars, which would provide fresh support for the MO model. At the same time, the inversion of the direction of polarization (PL and PR in Fig. \ref{fig:fig4}) can be explained simply by the MO model. The direction of the magnetic field has reversed, so the direction of circular polarization has also reversed.

Of course, even if this prediction is confirmed by subsequent investigations, it only adds evidence favoring the MO model but does not imply a complete substantiation for this model. If additional assumptions can be added, the lighthouse model may also be viable in explaining the results. Therefore, the decisive method remains that proposed by \citet{Liang14}.

\section{Discussion}
\citet{Desvignes19} reported fascinating results from their study of PSR J1906+0746 and provided staunch support for the lighthouse model. However, they modeled the system using at least seven parameters, which is inevitably worrisome. Using too many parameters sometimes reduces the credibility of the fitted results.

We acknowledge that the MO model is nascent. Even if correct, there are still many issues that need further research. We present below a few unsolved problems and preliminary thoughts.

Under the MO model framework, the magnetic field, electric field, and electron speed distribution are highly ordered, which creates conditions most favorable for coherent radiation. However, why the radiation occurs only when the electric field is strongest and the magnetic field is near its weakest remains unsolved.

The radiation mechanism can no longer be explained simply by cyclotron or synchrotron radiation because only the Lorentz force affects the electrons in these two processes. In the MO model, besides the Lorentz force, the induced electric field can force electrons to relativistic speeds within a few milliseconds. Therefore, it is appropriate to consider this radiation process as completely novel. Does this impact the discussions above and the prediction made in Section 4? This prediction was formed after considering this electric field, and the emergence of this alternating electric field and the novel radiation process is a necessary condition for the $90\degr$ phase lag. Therefore, the new radiation process does not impact this prediction. The addition of this electric field also does not impact our discussion of the polarization given above because the polarization is determined by the trajectory plane of the electron motion, and the added electric field does not change this plane (here the equatorial plane). The addition of an alternating electric field does affect for example the radio frequency spectrum, pulse phase, breathing-like changes in the radiation cone, and coherence characteristics.

The magnetic field curves reported contain both alternating and persistent components \citep{Berger08, Babcock51}. The zero point of the magnetic field in Fig. \ref{fig:fig4} only corresponds to the zero point of the alternating component, with no consideration of the persistent component. Our preliminary thought is that the alternating component may represent the dipole magnetic field of the star, whereas the persistent component may be a mixing effect generated by local magnetic fields on the star's surface. Periodic coherent radiation may occur at a distance from the surface of the star where the dipole magnetic field dominates.

The MO model admits magnetic periods of some stars to be close to millisecond level. If correct, we must abandon fossil magnetic field models and various dynamo models that rely on internal turbulence. The new dynamo model must generate purely electromagnetic oscillations driven by rotational energy. The period depends on the distributed capacitance, inductance, and rotational speed of the star. The smaller the volume and the faster the star rotates, the higher the oscillation frequency. Other influencing factors include temperature, stellar wind flux, and relativistic effects, among others.

If the MO model is ultimately confirmed, future pulsars might be divided into four types: hour pulsar (non-compact star), minute pulsar (white dwarf), normal pulsar (neutron star), and millisecond pulsar (recycled pulsar).

%If intelligent life outside the solar system observes our Sun, they will see only a point light source because of distance. With our current equipment, they will record various curves such as magnetic field intensity, solar irradiance, radio brightness, and X-ray brightness of the Sun. However, all this information will not establish the rotation period of the Sun as 25 days near the equator and 34 days near the poles. They will see a cycle of 11 years in each curve. They may also mistake the rotation period of the Sun as 11 years. We also encounter similar difficulties when observing other stars. Perhaps all the periodic signals of stars we have obtained are their magnetic periods or the radial respiratory periods and have nothing to do with the rotation period.

\section*{Acknowledgements}
We would like to thank the anonymous reviewer for the helpful comments. We thank Richard Haase, PhD, from Liwen Bianji (Edanz) (https://www.liwenbianji.cn) for editing the English text of a draft of this manuscript.
\section*{Data availability}
All data sources have been listed in the text.

\end{document}